# Changes from Classical Statistics to Modern Statistics and Data Science

Kai Zhang[1], Shan Liu[2] and Momiao Xiong[3,*]

[1] Department of Environmental Health Sciences, School of Public Health, University at Albany, State University of New York, Rensselaer, NY 12144, USA

[2] Department of Data Science, School of Math & Computer Science, Jiangxi Technology & Science Normal University, Nanchang, 330038, China

[3] Department of Biostatistics and Data Science, School of Public Health, The University of Texas Health Science Center at Houston, Houston, TX 77030, USA

**Key words:** Artificial intelligence, data science, manifold, geometric learning, diffusion model and composite diffusion

[*]Address for correspondence and reprints: Dr. Momiao Xiong, Department of Biostatistics and Data Science, School of Public Health, The University of Texas Health Science Center at Houston, P.O. Box 20186, Houston, Texas 77030, (Phone): 713-259-2371, (Fax): 713-500-0900, E-mail: Momiao.xiong@gmail.com.

**Abstract**

A coordinate system is a foundation for every quantitative science, engineering, health and medicine. Classical physics, mathematics and statistics are based on the Cartesian coordinate system. The classical probability theory and hypothesis testing theory can only be applied to Euclidean data. However, modern data in the real world are from directional statistics, natural language processing, translation, speech recognition, mathematical formulas, computer programs, social networks, transportation networks, sensor networks, computer visions, automations, biomedical and biomolecular measurements. The Euclidean assumption is not appropriate for non-Euclidean data. This perspective addresses the urgent need to overcome those fundamental limitations and encourages extensions of classical probability theory, hypothesis testing and regression, Brownian motion, diffusion equations and stochastic differential equations from Euclidean space to non-Euclidean space. Artificial intelligence such as natural language processing, translation, computer vision, graphical neural networks, manyfold regression, probability and inference theory, manifold learning, graph neural networks, compositional diffusion models for automatically compositional generations of concepts and demystifying machine-learning systems, has been rapidly developed. Differential manifold theory is the mathematic foundations of deep learning and data science as well. We urgently need to shift the paradigm for data analysis from the classical Euclidean data analysis to both Euclidean and non-Euclidean data analysis and develop more and more innovative methods for describing, estimating and inferring "non-Euclidean" geometries of modern real datasets. A general framework for integrated analysis of both Euclidean and non-Euclidean data, composite AI, decision intelligence and edge AI provide powerful innovative ideas and strategies for fundamentally advancing AI. We are expected to marry statistics with AI, develop a unified theory of modern statistics and drive next generation of AI and data science.

## Introduction

Statistics is to study the state of complex systems. It collects, summarizes, explores and analyzes the data from complex systems, uncovers the mechanism underlying the complex systems and predicts their future state. In history, statistics arose from the population survey (16$^{th}$ and 17$^{th}$ centuries), gambling (18$^{th}$ century), scientific data analysis (19$^{th}$ century), agriculture (19$^{th}$ and 20$^{th}$ centuries) and the development of probability theory, which put statistics on a firm theoretical basis ( Wikipedia, 2022).

In the last century, statistician including Ronald Aylmer Fisher, Karl Pearson, and Jerzy Neyman, laid the foundation of statistics. It includes developing the distributions of some statistics, e.g., normal distribution, the student's t distribution, F distribution, $\chi^2$ distribution, correlation coefficient and the multiple correlation coefficient, maximum likelihood as a major method for parameter estimation, the theory of estimation and hypothesis testing, linear and generalized linear models, variance analysis, large sample theory, classification and cluster analysis (Krishnan 2022). Although the classical statistics achieves great success in both theory and real-world data analysis, it shows several limitations which now seriously hamper its development and applications.

The first, the data all the classical statistical methods analyze are generated in Euclidean space. However, the current data such as natural language data, image data, video and audio data, cognitive data, mathematical formulas, and all human intelligence data are non-Euclidean data. The second, widely used models in statistics are linear or generalized linear models. However, nonlinear models are ubiquitous in the real world. The third, classical statistics deal with "out of distribution" problem via large sample theory. However, the distribution may be shifted in the real world, large sample theory cannot cover shifted distributions. The fourth, the widely used statistical methods for analyzing the relationships between random variables are association analysis. However, the association analysis are difficult to detect the mechanisms underlying the data generating processes. The fifth, statistical methods lack methods for studying individuals. The sixth, statistics are difficult to study qualitative characters. The seventh, the most classical statistics are lack of creative mechanisms and generative models.

**AI is changing the paradigm of data collection and analysis**

Artificial Intelligence (AI) is a technology which tries to simulate human reasoning, learning and thinking. AI is now used everywhere, from language translation, robotics, self-driving cars to our daily lives. AI is changing the paradigm of data collection and analysis. AI has five remarkable features.

*A coordinate system is a foundation for every quantitative science*

First and most important feature is the embedding of the subject, which is its semantic representation. A coordinate system is a foundation for every quantitative science, engineering, health and medicine. Without the Cartesian coordinate system, there would be no modern mathematics, physics and in general, science. Cartesian coordinate system transforms the various geometric shapes, physical quantities and sensor measured signals into the real numbers. However, many data types in AI applications such as information, words, images, sounds, traffic flows, DNA, proteins, genomic data, molecules, mathematic formulas, graphics cognitive memory and reasoning cannot be measured in the Cartesian coordinate system. These data are measured in the information coordinate systems (Kowsher et al. 2022; Frome et al. 2013; Levy et al. 2020; Akbari et al. 2021; Ji et al. 2021; Xu 2020; Wang et al. 2022; Krstovski and Blei 2018). To use the mathematical and statistical tools for analyzing these widely used data, we need to map them from the information coordinate system to the vector space while preserving the sematic meaning of the data in the original information coordinate systems. This vector space is also often called latent space. The variables in the vector space are called the embeddings of the subjects. The subjects in the information coordinate system are called human semantics (Schwalbe 2022). The embedding of subject in the vector space reflects the semantic representation in the human brain (Hollenstein et al. 2019; Ansarinia et al. 2022). A key to the success of application of multimodal AI to cognitive science is joint embeddings or representations of video, audio, image and natural language processing (Akbari et al. 2021).

*Universal approximation of the neural networks*

Second feature of the AI is the universal approximation of the neural networks (Kutyniok 2022). Assume that there is a training dataset $\{x_i, y_i\}_{i=1}^{m}$ and there exists a target output function $g: R^d \to R$ such that $y_i = g(x_i)$. Assume that we use a function $f: R^d \to R$ to approximate $g$. Define the risk of function approximation as

$$R(f) = \int_\mathcal{X} (f(x) - g(x))^2 dx . \tag{1}$$

The error between the trained neural network $\Phi^{(0)}$ and the target output function $g$ is bounded by (Kutyniok 2022):

$$R(\Phi^{(0)}) \leq \left[\hat{R}(\Phi^{(0)}) - \inf_{\Phi \in NN_\theta} \hat{R}(\Phi)\right] + 2 \sup_{\Phi \in NN_\theta} |R(\Phi) - \hat{R}(\Phi)| + \inf_{\Phi \in NN_\theta} R(\Phi) , \tag{2}$$

where

$\hat{R}(\Phi) = \frac{1}{m}\sum_{i=1}^{m}(\Phi(x_i) - y_i)^2$, $NN_\theta$ denotes a neural network with the parameter $\theta$.

The risk upper bound consists of three errors: (1) the approximation error defined as $\inf_{\Phi \in NN_\theta} R(\Phi)$, (2) the optimization error, defined as $\hat{R}(\Phi^{(0)}) - \inf_{\Phi \in NN_\theta} \hat{R}(\Phi)$ and (3) the generalization error, defined as $\sup_{\Phi \in NN_\theta} |R(\Phi) - \hat{R}(\Phi)|$.

Classical results for the universal approximation theorem focused on shallow neural networks (Pinks 1999). After the development of deep learning, researchers showed that the functions can be approximated by deep and narrow neural networks (Lin et al. 2017). Instead of bounded depth, the universal approximation theorem is also extended to the neural network of bounded width and arbitrary depth where the minimum width required for the universal approximation of the $L^p$ function is $\max(d_x + 1, d_y)$, $d_x$ is the input dimension and $d_y$ is the output dimension (Park et al. 2021). Exactly theoretically deriving the optimal structure of the neural networks is difficult or even impossible.

*Learnable function*

The AI develops a concept of learnable function (Wang 2021). The structure and parameters of the neural networks are automatically learned from the data via some algorithms. The learnable function concept dramatically improves the performance of neural networks as a powerful tool for prediction and regression.

*Pre-training, self-supervised learning, transfer learning and foundational models*

The third feature is pre-training, self-supervised learning, transfer learning and foundational models (Bommasani et al. 2021; Du et al. 2022). Pre-training, self-supervised learning and transfer learning are the focus of natural language processing (NLP), computer vision (CV) and multimodal learning. A major tool for dealing with samples in statistics is large sample theory which assumes a single population distribution. However, a population consists of multiple subpopulations. Each subpopulation has its own distribution. Sine in practice, the number of samples is limited. The data we observe come from one or a few specific subpopulations. The asymptotic results are unable to reveal the distributions of multiple subpopulations. This will deteriorate the performance of prediction using classical statistical methods.

To overcome such limitations, pre-training addresses three important issues. The first issue is how the pre-training imitates human to acquire background knowledge of how the world works. The second issue is what model with a large size can be used to fit the data. The third issue is how to generalize the model "out of distribution".

To address the first issue, the self-supervised learning mask out some tokens and then predict the masked context giving the unmasked visible text, which leads to acquiring the semantics of the data (Wei et al. 2021).

*Large models generate the intelligence pattern*

A large model is used to address the second issue. The large models are referred to as more than tens of gigabytes in size and trained on huge amounts of data. The recently released Pathway Language Model (PaLM) uses 540 B parameters and 780 billion tokens to train on 6144 TPU v4 chips (Chowdhery et al. 2022). The large models dramatically improve the performance across various tasks while using zero-short or few short learning.

Basic ideas behind the power of the large models for cognitive and creative tasks is unclear. The efficiency of training large models depends on the ratio of the model parameters over the number of training tokens (Hoffmann et al. 2022). If we think of each cognitive task and creative task performed in a subpopulation, and language and reasoning pattern underlying the cognitive and creative tasks as their distributions, then increasing the model size will increase the capability of the large model storing the distributions of the subpopulation which will lead to improving the performance of the large model's intelligence. The size, architecture, and parameters of the neural network generate the intelligence pattern which will provide tools for solving intelligent tasks. The simple structure of statistics will not have the such capability of doing such intelligent tasks.

The pre-trained single large model, surprisingly, can be used for tasks including translation, document summarization, question answering, computer vision, speech recognition, DNA generation, genomic analysis, programming, complex mathematic equation solving, and automated theorem proving (Noorbakhsh et al. 2021; Shen et al. 2021; Davies et al. 2021; Polu and Sutskever 2020). The role of different transformers and large models including universe transformer, is to change the architecture and parameter space of the neural networks, which in turn, change the capacity to work on the logic reasoning of the neural networks (Dehghani et al. 2019).

*Large scale machine learning is a major method to address the issue on generalizing to "out-of-distribution"*

Large scale machine learning and transfer learning are major methods to address the third issue on generalizing to "out-of-distribution". Improvements in pre-training would also lead to the improved accuracy for most downstream tasks (Abnar et al. 2022). In other words, improving the performance of pre-training on massive corpus would enable us to finish many downstream tasks at almost no costs. However, it is also reported that when increasing the accuracy of upstream tasks reaches a certain point, the accuracy of a downstream task becomes saturated (Abnar et al. 2021). After the accuracy of the downstream task is saturated, the strategies for continuing transfer learning includes data augmentation, developing causality-based models and designing efficient learning strategies such as invariant learning and adversarial training (Li et al. 2022). *Classical statistics are difficult to offer basic ideas and techniques contained in large models, massive samples, pre-training and transfer learning.*

**Causal Inference is a powerful reasoning tool and a natural extension of correlation**

*Correlation is to measure dependent relationships between variables. Causation is to measure the effect of intervention*

Correlation analysis and regression are basic concepts in statistics. Correlation is to measure dependent relationships between variables. Causation is to measure the effect of intervention. Causation is intuitively understood as that variable A is enforced to make changes, it will lead to change of variable B. Variable A is called the cause and variable B is called the effect. Causation implies correlation, but correlation does not conclude causation. Unobserved confounders often create spurious correlations, which leads to biased estimation of causation and blurs statistical analysis (Shapiro 2008; Yang et al. 2021). Human mind often likes to subjectively link dependent patterns as causal patterns (Madhavan 2019). We can observe many possibilities for correlation, but not for causation:

(1) Variable A causes variable B implies the correlation between variables A and B.
(2) Oppositely, B causes A also implies the correlation between A and B.
(3) Confounder C causes A and B implies the correlation between A and B.
(4) Third variable D is involved. A causes B as long as D event happens.

(5) Cause transitivity. A causes B, which leads B to cause C. It implies the correlation between A and C, but not implies that A causes C (Cai et al. 2019).

Causation is often defined as an intervention in an experiment where terminologies of system and control theory can be utilized. The intervention action is called a cause and the outcome generated by the intervention is called an effect. Since the experiments are often expensive, time-consuming, infeasible, and even unethical, we often use observational data to approximate experiments. To do this, we need to define a system. A causal model defined at the unit level consists of the following components (Pearl 2000):

Exogenous variables $X$ (causes), representing factors outside the system, which affect the endogenous variables $Y$ (effects) within the system where we assume that the probability distribution $P(X)$ of cause is independent of the conditional distribution $P(Y|X)$, given $X$. A special case is an additive noise model (ANM):

$$Y = f(X) + e,$$

where $e$ is a noise and is independent of cause $X$.

Intervention can be defined via a mathematical operator called $do(X)$ (Pearl 2012). The do-operator models physical intervention by replacing the cause with a constant $X = x$, denoted by $P(Y|do(X = x))$. Therefore, for the above ANM, we have

$$P(Y|do(X = x)) = P(e = y - f(X)).$$

To test whether $X$ causes $Y$ or not, we can fit the ANM model and then test the independence of $X$ and $e$. If $X \to Y$, then we can prove $P(Y|do(X = x)) = P(Y|X)$, i.e, the intervention distribution $P(Y|do(X = x))$ is equal to the conditional distribution $P(Y|X)$ in statistics.

However, if $Y \rightarrow X$, then $P(Y|do(X = x)) = P(Y)$, in this case, the intervention distribution $P(Y|do(X = x))$ is equal to the marginal distribution $P(Y)$. In other words, if the causal model is identifiable, then the intervention distribution $P(Y|do(X = x))$ can be reduced to the statistical conditional distribution $P(Y|X)$. This shows that the causal inference is a natural extension of correlation in statistics.

*A basic framework for causal inference is that we "repeatedly apply the rules of do-calculus until the final expression no longer contains a do-operator"*

For more than three variables, Pearl (1995, 2012) developed three do-calculus rules (Rule 1: Insertion/deletion of observations, Rule 2: action/observation exchange, Rule 3: Insertion/deletion of actions). A basic framework for causal inference is that we "repeatedly apply the rules of do-calculus until the final expression no longer contains a do-operator" ( Pearl 2012). We start with "do" intervention distribution and end with the do-free conditional distribution in statistics if the causal model is identifiable. It shows that if the do-operators cannot be removed by repeatedly applying the do-calculus rules, then the model is unidentified.

We present two examples to illustrate how to use do-calculus for transforming causa-effect estimation to a series of conditional distribution calculations. The first example is to learn individual-level causal effects in the presence of unmeasured confounders from observational data. Louizos et al. (2017) considered the latent confounders, covariates, treatment and outcome. They showed that the calculation of treatment effect can be reduced to integral of the conditional probability of outcome, given treatments, covariates and confounders via repeatedly applying do-calculus rules, and design a variational autoencoder that recovers the joint distribution of the outcome, treatment, observed covariates and hidden confounders from the observations of the outcomes, treatment and covariates.

The second example is causal attention for vision-language tasks (Yang et al. 2021). The ever-elusive confounding effects in the attention-based vision-language models make unrequired bias that misleads us to attend the spurious correlations in the training data. They used front-door adjustments, In-sample sampling and Cross-sample sampling to reduce the intervention distribution to the summation-product of the statistical conditional distribution. They further developed In-Sample attention and Cross-Sample attention to form causal attention for vision-language tasks.

This convinces us that although causal inference is different from associations, causal inference from observational data is indeed, a natural extension of statistical association analysis.

## Breakthrough and emergent abilities of large language models

*Language models demonstrate not only smoothing quantitative improvement, but also unpredictable abrupt emergent abilities*

Natural language process (NLP) tasks include text classification which takes input text to a deep neural network-based encoder to predict the class label (Howard and Ruder, 2018), matching that predicts the sematic relevance of two texts (Ghosal et al. 2022), sequence labeling (He et al. 2020), text summarization (Zhang e al. 2022), machine translation and dialogue (Gorcia and First, 2022), commonsense reasoning (Bhargava and Ng, 2022), and language model (LM) which estimates the probability of sequence words, given their contents (Zhou et al. 2020). A new paradigm for NLP is to pretrain language models with prompts, which combine diverse pre-training objectives together and universally work well across different tasks (Sun et al. 2022).

The performance of LM depends on the amount of available data, computing power and the number of parameters. Large generative language models have recently grown exponentially. These models include 20B UL2 (Tay et al. 2022), 175B GPT3 (Brown et al. 2020), 176B Bloom (BigScience Worksop, 2022), 280B Gropher (Rae et al. 2021), 137B LaMDA (Thoppilen et al. 2021), 137B FLAN (Wei et al. 2022), 178B Jurassic-1 Jumbo (Reed et al. 2021), 200B PanGu (Zeng et al. 2021), 204B HypercLoVA (Kim et al. 2021), 246B Yuan 1.0 (Wu et al. 2021), 530B Megatron Turing NLG (Turing et al. 2022), 540B psthways Languaage Model (PaLM) (Chowdhery et al. 2022).

When the sizes of language generative models increase, we observe several remarkable features (Granguli et al. 2022; Srivastava et al. 2022; Wei et al. 2022).

1. Language modeling performance depends most strongly on scales (the number of model parameters, the size of available dataset and the amount of computing used for training), but weakly on depth and width of the architecture of the neural networks (Kaplan et al. 2020).
2. General capability and performance of language models smoothly improve with increasing number of model parameters, compute and data sizes.
3. Language models demonstrate not only smoothing quantitative improvement, but also unpredictable abrupt emergent abilities (Wei et al. 2022). A total of 444 authors from 132 institutions design the beyond imitation game bench-mark (BIG-BENCH) that consists of 204 tasks (Srivastava et al. 2022). Task topics include problems from linguistics, mathematics, common sense reasoning, health and biology, physics, social bias, programming coding, etc). BIG-BENCH is tested by three language models: Open AI's GPT models, Google internal transformer architectures, and Switch-style sparse transformers. The results show the "breakthrough" behavior of the language models where the performances change from quantitative improvement to new emerging qualitative capabilities. The abrupt changes in the behaviors of the large language models may have large beneficial or harmful impact on the society.

*Abrupt changes in the performance curve can be detected by the symmetrized KL distance*

There are clear evidence that the cross-entropy loss of the language generative models such as generative imaging modeling, video modeling, mathematical problem solving that requires some level of reasoning, decreases as the number of parameters, the sizes of available training data and compute budget increase (Henighan et al. 2020). The loss function can be empirically approximated by

$$Loss = S(True) + D_{KL}(True|model),$$

where $S(True)$ represents an entropy of the true model, and is often called irreducible loss and $D_{KL}$ represents the Kullback-Leibler (KL) distance between the ground truth and model distribution and is often called the reducible loss. The entropy here is defined for the curves of performance of language model (Balestrin et al 2009). The entropy of the curves measures the complexity of the curves and dynamic system. If the curve is a straight line, then its entropy is 1, or the dynamic system is linear, then again, the entropy of the linear dynamic system is 1. The entropy grows with nonlinearity and irregularity of the curve. The performance curve of the large language model can be modeled as a piecewise stationary dynamic system. Abrupt changes in the performance curve can be detected by the symmetrized KL distance (Last and Shumway 2008).

Recently, machine learning based model free, data driven methods for predicting abrupt changes or catastrophic collapse due to parameter drift without relying on a model have been developed (Kong et al. 2021; Patel and Ott, 2022). These model-free machine learning methods can be used to predict tipping points and abrupt changes of the performance of large language models and investigate how the continuously quantitative changes lead to discontinuously qualitative changes of the performance of large language models.

**Probability distribution and hypothesis testing in manifolds**

*Probability theory of Kolmogorov and Fisher and Neyman theory of hypothesis testing are developed for Euclidean space*

Andrey Nikolaevich Kolmogorov developed the foundation of probability distribution. He was born in Tambov in 1903 and enrolled at Moscow State University in 1920. He studied set theory and the theory of Fourier series during his undergraduate years and right after his graduation. Then, with the help of his knowledge in set theory, He published his famous paper "On the principle of the excluded middle" in 1925 and in published the book "Foundations of the Theory of Probability" in 1933, axiomatizing the probability theory and defining probability distribution in **Euclidean space** (Sack 2018).

During the same period of time, in 1920, Ronald Fisher proposed to use p value for hypothesis testing and Jerzy Neyman and Egon Pearson directly developed the theory of hypothesis testing. The observed data and testing regions are in Euclidean space and the probability distributions for test statistics are defined in Euclidean space. Therefore, the classical probability theory and hypothesis testing theory can only be applied to Euclidean data (Biau et al. 2010).

*Manifold probability theory and hypothesis testing emerge*

However, the Euclidean assumption is not appropriate for non-Euclidean data (Galaz-Garcia et al. 2022). Typical examples of non-Euclidean data include directional statistics, graphics, mathematical formulas, computer programs, social networks, transportation networks, sensor networks, computer visions, automations, biomedical and biomolecular measurements, and latent variable models with the parameters in non-Euclidean spaces. Unlike classical probability theory where the values random variables take are in Euclidean space, modern probability theory also considers manifold-valued random variables and constructs a probabilized space with manifold-valued outcomes (Pennec 2004). Probability theory on Riemannian manifolds is divided into two intrinsic and extrinsic methodology. Pennec (2006) gives a general definition of probability measures on Riemannian manifolds.

Many data in the real world are high dimensional. Numerous methods for analyzing high dimensional data are based on based on the assumption that data in high dimensional space tend to lie near a low dimensional manifold. Hypothesis test theory defines hypothesis in manifold, decision regions are defined in manifolds and probability distribution of the test statistics is calculated based on the probability theory on Riemannian manifolds (Fefferman et al. 2016; Wang et al. 2022).

*Manifold regression*

There is an increasing interest in functional regression on manifolds (Torres-Signes et al. 2022; Guo et al. 2020; Lin and Yao 2021; Chen et al. 2022; Hacquard et al. 2022). Manifold regression can be divided into three categories. First category is manifold-valued response regression where a responsible variable is manifold-valued and predictor can be either categorical or real-valued data (Random forest regression for manifold-valued regression, and geodesic distance and multivariate distance matric regression for manifold-valued responsible variant) (Tsagkrasoulis and Montana 208; Ryan et al. 2020). Second category is functional regression with response in Euclidean and prediction in manifolds. The manifold regression in second category often uses Riemannian functional principal component analysis or the Laplace-Beltrami eigenbasis (Dai and Muller 2018; Hacquard et al. 2022). Third category is manyfold regression with both manifold-valued response and manifold-valued predictors (Cenedese et al. 2022). However, methods for third category of manifold regression have not been well developed.

**Challenge of manifold learning and computational biology**

*Manifold learning with symmetry and invariance as its core elements is " geometric foundations of deep learning"*

Traditional statistics and machine learning have mainly focused on the analysis of data in Euclidean vector spaces. Many data from social networks, biomedical and molecular measurements, transportation networks and sensor networks which AI studies are non-Euclidean data with manifold and topological structures (Gilmer et al. 2017; Nguyen et al. 2022; Cao et al. 2022; ICLR 2022 launched "challenge for computational geometry & Toplogy: design and results" ; Mayers et al. 2022). Manifold learning with symmetry and invariance as its core elements is " geometric foundations of deep learning" (Bronstein, 2021). Manifold learning effectively combines deep learning and non-Euclidean data consisting of manifolds and graphs.

*The classical calculus and Euclidean geometry are the mathematical foundations of the classical mechanics*

The classical calculus and Euclidean geometry are the mathematical foundations of the classical mechanics. Developments in the study of electricity and magnetism gave rises to the development of electromagnetic theory and the discovery of Maxwell's equations form the foundations of classical physics which describe and explain the motion of ordinary physical objects in the Euclidean space.

*Differential manifold is foundation of modern physics*

In 1915, Einstein published his general relativity theory (Jones and Robins, 2022). He defined a set of field equations in which gravity was described as the bending of space-time geometry. Einstein assumed that an object with mass must curve the space-time field. British astronomer Arthur Edditon observed the deflection of light by the sun during an eclipse in 1919 (Jones and Robbins, 2022), which matched a prediction by general relativity theory. The foundation of general relativity theory was changed from Euclidean geometry to differential geometry – a non-Euclidean geometry.

"Quantum differential manifold" plays the role of the carrier space in quantum mechanics (Ciaglia et al. 2019). Application of differential manifold theory to quantum physics includes Lagrangian and Hamiltonian formalisms, canonical quantization, gauge theory and diffusion theory. These form some of the central concepts in modern theoretic physics (akahera, 2003). In summary, the differential manifold which works on non-Euclidean data is one of the mathematical foundations of modern physics.

*Differential manifold is the mathematical foundation of deep learning*

Most tasks of AI deal with non-Euclidean data. AI is another area in which differential manifold works. The application of manifold to AI is referred to as manifold learning or geometric learning (Lei et al. 2018). Manifold learning consists of two types of problems: (1) understanding the mathematical foundation of deep learning and uncovering its mystery and (2) developing algorithms to learn the manifold data. The basic assumption for manifold learning is that high dimensional data are embedded in a nonlinear low dimensional manifold. A framework for manifold learning is to learn the manifold structure and the probability distribution of it from the data. A basic block of deep learning is a multilayer feedforward neural network with piecewise linear geometry. Its affine and linear transformation induces a set of polyhedral complexs which approximate the manifold in the data and implicitly capture a local coordinate system for this manifold (Lei et al. 2018; Lei et al. 2020). Therefore, mathematic foundation of deep learning is differential manifold.

*Manifold leaning extends the traditional real-value-based optimization to manifold-valued optimization*

Manifold learning involves every aspect of the data science. Manifold leaning extends the traditional real-value-based optimization to manifold-valued optimization (Shustin et al. 2022; Boumal 2022; Duruisseaux and Lesk 2022; Han et al. 2022; Bu and Chang 2022; Jin and Sra 2022; Hu et al. 2019). Manifold optimization is to find the desired optimum that is constrained to a small manifold. First-order and second-order optimality conditions for real-valued optimization are extended to manifold optimization. Many methods for manifold optimization such as the Riemannian gradient method and Riemannian Newton's method are developed.

*Manifold alignment for integration of multi-modal data*

Many data analysis deal with multiple datasets from different (heterogenous) domains. Data from different domains may come from the same samples. For example, single cell multi-omics data are measured from the same sample (Cao et al. 2021). Alternatively, although different datasets may not come from the same samples, they indeed, come from the similar subjects, e.g., words and sentences of different languages. Manifold alignment is to find the shared low dimensional manifold of multiple datasets (Chen et al. 2022). Manifold alignment can be classified into supervised (Schneider et al. 2014), semi-supervised (Aziz et al. 2019; Hong et al. 2019) and unsupervised manifold alignment (Xu et al. 2022; Singh et al. 2022; Yin et al. 2022). Supervised manifold alignment requires that one-to-one correspondence between domains of entire dataset is known. In many cases, this may not be realistic. Realty, methods for manifold alignment have focused on the semi-supervise and unsupervised alignment. These methods include diffusion transport alignment that exploits prior correspondences between only a few samples to align the domain (Duque et al. 2022), Riemann geometry of symmetric positive-definite matrices for alignment which exploit Laplace-Beltrami operator (Shnitzer et al. 2022), and topological autoencoders for unsupervised manifold alignment which first discover latent manifold representation of each modality separately, and then use topology preserving generative adversarial network (GAN) to align these latent representations into a common manifold space (Singh et al. 2022).

*Manifold learning has primarily focused on graph learning*

Graph is one of major discrete manifolds. Manifold learning has primarily focused on the field of graph neural networks (GNNs). Node-based message-passing mechanism generates the first generation of GNNs, which are often referred to as massage passing networks (MPNNs) (Balcilar et al. 2021). Message passing is a form of signal diffusion (Chamberlain et al. 2021). MPNNs exchange information between the neighboring nodes in each layer (Gilmer et al. 2017; Gasteiger et al. 2022), and diffusion information on the graph using a trainable nonlinear function (Topping et al. 2022). MPNNs lead to the development of various GNNs. Despite their successful applications in deep learning, GNNs following the massage passing paradigm, have the limited expressive power (Xu et al. 2019), over-smoothing (Oono and Suzuki, 2020) and over-squashing (Alon and Yahan, 2021). To overcome these limitations, differential geometry is used to study the bottleneck and over-squashing phenomena in MPNNs. It has been shown that negatively curved edges cause the formation of bottlenecks and over-squashing. Stochastic discrete Ricciflow is used to alleviate the over-squashing (Topping et al. 2022).

*Extend diffusion process from Euclidean space to non-Euclidean space*

Massage passing in graph is the generalization of the diffusion process to graphs, a special case of manifold. Exchanging information in GNNs is conceptually equivalent to diffusion where the base space is the graph, diffusion occurs along edges, and the analogy of the spatial derivatives is the differences between neighboring node features (Gilmer et al. 2017; Bronstein et al. 2021; Chamberlain et al. 2021). Information diffusion on graphs is intimately connected to differential equations (partial differential equations (PDEs) and stochastic differential equations (SDEs)). Both PDEs and SDEs can be used to design GNN architectures and develop a broad class of new algorithms of GNNs (Chamberlin et al. 2021; Croitoru et al. 2022; Eliasif et al. 2022; Gao et al. 2022; Jo et al. 2022; Yang et al. 2022; Song et al. 2021; Bortoliet al. 2022; Dockhorn et al. 2022; Dhariwal and Nichol, 2021).

*Both PDEs and SDEs can be used to design GNN architectures and develop a broad class of new algorithms of GNNs*

It showed that the explicit single-step Euler scheme for the PDEs is equivalent to the standard GNNs, Runge-Kutta scheme performs significantly better than the single-step Euler scheme, and implicit schemes are equivalent to larger multi-hop diffusion operator which ensures that information can exchange with the remote nodes (Chamberlain et al. 2021). Discrete PDEs on graph can be exploited to design very deep GNNs upto tens of layers, which leads to overcoming the problems of over smoothing and bottlenecks.

*Diffusion models provide a powerful tool for deep generative models*

Diffusion models provide a powerful tool for deep generative models. Diffusion models have found their wide applications to image generation, image super-resolution, image inpainting, image editing, image-to-image translation, understanding social networks, drug design and computer program synthesis (Cao etal. 2022; Brookschmidt et al. 2019).

Diffusion models can be classified into three sub-categories. The first sub-category is denoising diffusion probability model (DPMMs) which are latent models employing latent variables to estimate the probability distribution. DDPMs are a special case of variational autoencoders (VAEs). Its forward diffusion can be viewed as the encoding process, while its reverse diffusion stage can be taken as the decoding process in the VAE.

The second sub-category is noise conditioned score networks (NCSNs) which train a shared neural network through score matching, where the score function (defined as the gradient of the log density of the perturbed distribution) at different noise levels is estimated (Croitore et al. 2022; Song et al. 2021). The third sub-category of diffusion models is stochastic differential equation approach. SDEs can learn the underlying distribution of graphs and capture the permutation-invariance property of graphs (Jo et al. 2022).

*Application of SDEs to generate a graph can model the joint distribution of the nodes and edges*

In general, the generative models via SDEs unify the forward diffusion process that generates samples from noise by perturbing the data with gradually increasing noise and the reverse diffusion process that reverses the perturbation to gradually remove the noise into a single framework that implements the noise perturbation by SDEs. , The application of SDEs to generating a graph, can model the joint distribution of the nodes and edges (Jo et al. 2022).

*Diffusion-based generative AI can produce creative images and arts which can compete with humans*

Generative models including GANs, VAEs, normalizing flow models and one-short graph generating models, have been successful in generating images and graphs. However, it was reported that "diffusion models can achieve image sample quality superior to the current state-of-art generative models" (Dhariwal and Nichol, 2021). The methods underlying DALL, stable diffusion and other types of generative AI is a diffusion model. These generative AI can produce creative images and arts which can compete with humans (Big Think, 2022; Gordon 2022). The traditional statistical methods which lack intelligent and creative mechanisms underlying them, can never generate creative images and arts.

Human can learn new concepts from limited examples (Zhou and Lake 2021) and is able to compose complex concepts out of simpler ideas and knowledge, which lead to rapid learning and adoption of knowledge (Du et al. 2020; Huang et al. 2022; Lin et al. 2021; Liu et al. 2022). A popular method for composing concepts is a diffusion model. Specifically, diffusion models are interpreted as energy-based models where concepts modeled by the distribution of the data and energy function can be explicitly combined (Liu et al. 2022). Diffusion models with compositional operators can first learn different portions of the model, one at a time, then compose different concepts together using logical conjunction and negation operators. This implies that composite diffusion models allow to incrementally learn concepts in language and knowledge, and open a new direction for the creative generation of images, arts and continual learning (Li et al. 2021, 2022) to support dynamically growing number of tasks. Composite diffusion models elegantly formulate the concepts and explicitly composite diffusion models to generate an image, given a complex natural language text (Gordon 2022).

*Application examples*

Manifold learning has been successful applied to biomedicine. Typical examples are given as follows.

(1) protein sequence design

Using MPNNs or more general GNNs, and given a protein backbone structure of interests, we can find an amino acid sequence that will fold to this structure (Dauparas et al. 2022).

(2) Drug binding protein target prediction

Both ligand and receptor molecules are presented as K-nearest neighbors (K-NN). Graph matching networks and graph neural networks are combined to perform intra and inter neural graph message passing, and prediction of affinity of ligand binding receptors (Stark et al. 2022; Tabiana et al. 2022).

(3) Antibody binding prediction

A diffusion model is developed to jointly model the sequence and structure of the complimentary-determining regions between antibodies and antigens (Luo et al. 2022).

(4) Drug repurposing

Drug features from multiple heterogeneous networks and disease features from biomedical documents for drug repurposing are combined to find new potential candidates for approved drugs (Shahini et al. 2022; Jin et al. 2021).

(5) Integration of space and single cell omics data

Topology-preserving autoencoders are often used to obtain the latent representation of each modality separately and then deep generative models are used to align these latent representations into a common space (Singh et al. 2022).

**Conclusions**

There is an increasingly evident understanding that most data in AI cannot be modeled via simple Euclidean structures. The most data which AI works are non-Euclidean. Modern data in real world are from natural language processing, translation, speech recognition, mathematical formulas, computer programs, social networks, transportation networks, sensor networks, computer visions, automations, biomedical and biomolecular measurements, which have complex structures, nonlinear relations and interactions. The classical statistics have never worked on such types of data. These complex structures and relations can only be captured by "non-Euclidean" geometrics. Since we are only familiar with analyzing data in Euclidean space, the "non-Euclidean" data are embedded into Euclidean space for processing and analysis. The paradigm for data analysis is being shifted from classical Euclidean data analysis to both Euclidean and non-Euclidean data analysis. Numerous methods for language models, manifold learning, computational algebra and topology, graph representation and graph signal processing, and nonlinear analysis, have been developed to describe, estimate and infer "non-Euclidean" geometries of modern real datasets in AI (Wolf et al. 2022).

To meet the challenge of developing algorithms and computational tools for AI data analysis, composite AI, decision intelligence, and edge AI are emerging to be major innovations which will "hit mainstream adoption in two to five years" (Wiles 2022); Hernandez et al. 2022). Composite AI will automatically compose individual components in AI analysis into integrated intelligence activities to develop higher levels of knowledge representations and creativities. Composite AI learns from incomplete or partial datasets and enables us to establish intelligent machines that combine information into one comprehensive system (Actico 2021).

When applications of AI are facilitated, quality of AI improves, decision intelligence emerges. Decision intelligence aids governments, institutions and other organizations to make decisions, fully evaluate the risk consequences of their decisions and low the risk of their decisions.

Edge AI is decentralized computing that facilitates to make data-led decisions closer to where the data are generated (Actico 2021). Edge AI combines edge computing and AI. It implements AI algorithms and processing the data locally. Edge AI opens a new avenue to robust, fast and scalable AI systems across individuals, institutions and multiple industries.

A general framework for integrated analysis of both Euclidean and non-Euclidean data, composite AI, decision intelligence and edge AI provide powerful innovative ideas and strategies for fundamentally advancing AI. We are expected to marry statistics and AI to form a unified theory of modern statistics and to drive the next generation of AI and data science.

Dehghani, M., Gouws, S., Vinyals, O., Uszkoreit, J., Kaiser, L. (2019). Universal Transformers. arXiv:1807.03819.

Abnar, S., Dehghani, M., Neyshabur, B., Sedghi, H. (2021). Exploring the limits of large scale pre-training. arXiv:2110.02095.

Li, H., Wang, X., Zhang, Z., Zhu, W. (2022). Out-Of-Distribution generalization on graphs: A survey. arXiv:2202.07987.

Bellemare, M. F., Bloem, J. R., Noah Wexler, N. (2021). The Paper of how, estimating treatment effects using the front-door criterion. Ateliers méthodologiques de Montréal— Montréal Methods Workshop. Retrieved from https://semrasevi342192471.files.wordpress.com/2021/01/bellemare.pdf.

Madhavan, A. (2019). Correlation vs causation: Understand the difference for your product. Retrieved from https://amplitude.com/blog/causation-correlation.

Shapiro, S. (2008). Causation, bias and confounding: a hitchhiker's guide to the epidemiological galaxy. *J Fam Plann Reprod Health Care*, 34:261-264.

Pearl, J. (2009). *Causality: Models, Reasoning, and Inference*. Cambridge University Press. Second ed., New York.

Yang, X, Zhang, H., Qi, G., and Cai, J. (2021). Causal attention for vision-language tasks. arXiv:2103.03493.

Cai, R., Qiao, J., Zhang, K., Zhang, Z., Hao, Z. (2019). Causal discovery with cascade nonlinear additive noise models. arXiv:1905.09442.

Pearl, J. (2012). The Do-Calculus Revisited. arXiv:1210.4852.

Louizos C, Shalit U, Mooij J, Sontag D, Zemel R, Welling M. (2017). Causal effect inference with deep latent-variable models. arXiv:1705.08821.

Howard, J., Ruder, S. Universal language model fine-tuning for text classification. arXiv:1801.06146 (2018).

Ghosal, T, Saikh, T, Biswas, T, Ekbal, A, Bhattacharyya, P. (2022). Novelty detection: A perspective from natural language processing. *Computational Linguistics*, 48 (1): 77–117.

He, Z., Wang, Z., Wei , W., Feng, S., Mao, X., and Jiang, S. (2020). A survey on recent advances in sequence labeling from deep learning models. arXiv:2011.06727.

Zhang, Y., Ni, A., Mao, Z., Wu, CH., Zhu, C., Deb, B. et al., (2022). Summ^N: A multi-stage summarization framework for long input dialogues and documents. arXiv:2110.10150.

Garcia, X., First, O. (2022). Using natural language prompts for machine translation. arXiv:2202.11822.